\documentclass[apj]{emulateapj}
\usepackage{apjfonts}
\shorttitle{{\it Swift} monitoring of Cyg~X-2}
\shortauthors{Rykoff, Cackett \& Miller}

\begin{document}

\newcommand\swift{\emph{Swift}}
\newcommand\cyg{Cyg~X-2}
\newcommand\xte{XTE~J1817$-$330}
\newcommand\mdot{\odot{m}}
\newcommand\rxte{\emph{RXTE}}

\title{\emph{SWIFT} monitoring of Cygnus~X-2: investigating the NUV-X-ray connection}

\author{E.~S.~Rykoff\altaffilmark{1},
E.~M.~Cackett\altaffilmark{2,3}, and
J.~M.~Miller\altaffilmark{3}}
\email{erykoff@physics.ucsb.edu}
\altaffiltext{1}{TABASGO Fellow, Physics Department, University of California at Santa Barbara, 2233B Broida Hall, Santa Barbara, CA 93106, USA}
\altaffiltext{2}{{\it Chandra} Fellow}
\altaffiltext{3}{Department of Astronomy, University of Michigan, 500 Church St, Ann Arbor, MI 48109-1042, USA}

\begin{abstract}

  The neutron star X-ray binary (NSXRB) \cyg{} was observed by the \swift{}
  satellite 51 times over a 4 month period in 2008 with the XRT, UVOT, and BAT
  instruments.  During this campaign, we observed \cyg{} in all three branches
  of the Z track (horizontal, normal, and flaring branches).  We find that the
  NUV emission is uncorrelated with the soft X-ray flux detected with the XRT,
  and is anticorrelated with the BAT X-ray flux and the hard X-ray color.  The
  observed anticorrelation is inconsistent with simple models of reprocessing
  as the source of the NUV emission.  The anticorrelation may be a consequence
  of the high inclination angle of \cyg{}, where NUV emission is preferentially
  scattered by a corona that expands as the disk is radiatively heated.
  Alternatively, if the accretion disk thickens as \cyg{} goes down the normal
  branch toward the flaring branch, this may be able to explain the observed
  anticorrelation.  In these models the NUV emission may not be a good proxy for
  $\dot m$ in the system.  We also discuss the implications of using \swift/XRT
  to perform spectral modeling of the continuum emission of NSXRBs.

\end{abstract}


\section{Introduction}
\label{sec:intro}

The nature of near-ultraviolet (NUV) emission from neutron star low-mass X-ray
binaries (LMXBs) is not well understood.  The NUV emission provides essential
information in the broadband spectral energy distribution (SED) of the LMXB.
The NUV emission may be direct flux from the accretion disk; it may be hard
X-ray emission reprocessed by the accretion disk; or it may be dominated by jet
emission.  \citet{rfhbh06} have analyzed an ensemble of black hole LMXBs
(BHXRBs) and neutron star LMXBs (NSXRBs) to determine whether the NIR/optical
emission is more consistent with synchrotron (jet) emission or reprocessed
X-ray emission.  They point out that if the optical/NIR (and by extension the
NUV) spectrum is jet dominated, then it should be flat from the radio regime
through the optical, and $L_{\mathrm{Opt/NIR/NUV}} \propto L_X^{1.4}$.  One can
also make predictions for whether the NUV light is consistent with X-ray
emission reprocessed by the outer accretion disk. \citet{vpm94} show that under
simple geometric assumptions, reprocessed emission should be proportional to
$L_X^{0.5}\,a$, where $a$ is the orbital separation of the system.  Finally, if
the NUV emission is direct emission from the accretion disk, we might naively
expect the $L_{\mathrm{NUV}}$ to track $L_X$ directly.  In a study of an
ensemble of 13 NSXRBs observed over many orders of magnitude of X-ray
luminosity, \citet{rfhbh06} have shown that the optical/NIR emission is more
consistent with the prediction of reprocessed X-ray emission. Moreover,
\citet{hhohr06} have seen evidence for reprocessed emission in Type-I X-ray
bursts from the neutron star LMXB EXO~0748$-$676.

The \swift{} Gamma-Ray Burst Explorer~\citep{gcgmn04} is dedicated to the
discovery and follow-up of gamma-ray bursts.  It also is able to perform new
multi-wavelength studies of variable X-ray objects, revealing the nature of
accretion disks around compact objects.  On board \swift{} is the X-ray
Telescope~\citep[XRT,][]{bhnkw05}, an imaging CCD spectrometer with energy
coverage from 0.3-10 keV.  In addition,
the \swift{} UV/optical telescope \citep[UVOT,][]{rkmna05} can monitor the NUV
emission in NSXRBs, and the \swift{} Burst Alert Telescope
\citep[BAT,][]{bbcfg05b} can trace hard X-ray emission (14-300 keV) for bright
sources.

Using 21 short \swift{} monitoring observations of the black hole LMXB
XTE~J1817$-$330, \citet{rmst07} were able to show that the NUV flux tracks
closely with the 2-10 keV X-ray emission, with a best-fit slope of
$0.47\pm0.02$.  As the 2-10 keV X-ray emission is an effective proxy for the
hard X-ray emission detected by the BAT (see Section~\ref{sec:discussion}),
this is consistent with the hypothesis that the NUV emission is reprocessed
hard X-ray emission.  This was the first time that the NUV emission was
definitively shown to be reprocessed hard X-ray emission for a single black
hole LMXB source over a wide range of X-ray and NUV luminosities.  This is
consistent with the observations of an ensemble of black hole LMXBs in
\citet{rfhbh06}.  Thus, while it appears that reprocessing is important for
black hole LMXBs, the origin of the NUV emission in neutron star LMXBs is still
uncertain.

In order to investigate both the origin and evolution of UV and X-ray emission
in NSXRBs we have monitored the Z source \cyg{} approximately 50 times over a
4-month period using short $\sim$ 1~ksec observations with \swift.  \cyg{} is a
well known NSXRB, and has been used to define one of the types of Z
track observed (the so-called Cyg-like Z tracks).  It is well-suited to UV and
X-ray observations due to the low Galactic absorption in its direction
($2\times10^{21}$ cm$^{-2}$), thus it has been the subject of several previous
joint UV/X-ray monitoring campaigns~\citep{vrgvh90,hvedm90,vrbms03}.

During June and October 1988, \citet{vrgvh90} monitored \cyg{} with the
International Ultraviolet Explorer (IUE).  For seven of these IUE observations,
they also had simultaneous X-ray observations using \emph{Ginga}.  Over these 7
observations, \cyg{} was seen in all three branches of the Z track: horizontal
branch (HB), normal branch (NB) and flaring branch (FB). From these
observations \citet{vrgvh90} conclude that both the NUV continuum and line
emission increase monotonically along the Z track, with the least emission in
the HB, and the most in the FB.  They model the NUV emission as a combination
of reprocessed X-rays from the disk, along with a small (at most 20\%)
contribution from the X-ray heated surface of the companion star.  In addition,
they find no direct correlation between the NUV (continuum or line) flux and
X-ray (1-14 keV) flux from the observations. \citet{hvedm90} also discuss this
joint X-ray/NUV monitoring campaign of \cyg.  As the shape of the X-ray
spectrum changes over the Z track, they argue that the accretion disk is a
superior bolometer than our X-ray detectors because it is insensitive to
bandpass effects.  Assuming the NUV emission is dominated by reprocessed hard
X-ray emission, and thus the NUV emission tracks the total X-ray emission,
\citet{hvedm90} argue the NUV emission tracks the mass accretion rate,
$\dot{m}$.  Combined with the hint that the UV continuum increases along the Z
track, they conclude that the mass accretion rate increases along the Z track
from HB-NB-FB. Note, however, that \citet{chb06} suggest that the UV increase
from HB-NB-FB is not conclusive and does not imply an increase of mass
accretion rate.

Even after many decades of observations of the X-ray spectra of NSXRBs, there
is still not a clear consensus as to the spectral model to use, as a wide
variety of different spectral models often fit equally well
\citep[e.g.][]{barret01}. This has led to two classes of models being developed
to fit the spectra during the soft state: the {\it Eastern} model
\citep{mint89} comprised of a disk blackbody and a weakly Comptonized
blackbody; and the {\it Western} model \citep{wsp88} comprised of a
single-temperature blackbody from the boundary layer and Comptonized emission
from the disk. In contrast, the hard state is dominated by a hard component,
with the addition of a soft component which can either be a blackbody or a disk
blackbody~\citep[e.g.][]{bobds00}.  Given the spectral ambiguities, color-color
diagrams have been most frequently used to describe the behavior of NSXRBs.  In
fact, these sources are classed as either atoll or Z sources depending on the
shape they trace out on the color-color diagrams \citep{hv89}.

Nevertheless, a recent study by \citet{lrh07} using \rxte{} spectra of the
transient sources Aql~X$-$1 and 4U 1608$-$52 has provided progress on the
`correct' choice of spectral model.  These authors test all the commonly used
models and find that during the soft state the only model where the measured
temperature follows the luminosity as $L \propto T^4$ is the one comprised of
two thermal components, a disk blackbody and a single-temperature blackbody, in
addition to a broken power-law.  In this case, both the thermal components
follow $L\propto T^4$. This provides compelling physical motivation to use such
a model, which can be interpreted as emission from the accretion disk and from
a small boundary layer. However, the bandpass of \rxte/PCA is restricted to
above $\sim$3 keV, while the temperature of these components lies below 3 keV.
This can lead to inaccuracies in modeling, and these findings should be
confirmed by an instrument such as XRT with lower energy coverage.

How these sources progress around their tracks on the color-color diagram, and
what drives the changes, is still a matter of debate.  Recently, the transient
source XTE~J1701$-$462 has shown some unique properties which have important
ramifications.  It displayed both Z and atoll tracks, with the Z tracks (both
Cyg-like and Sco-like) occurring when the source was at its highest luminosity
and evolving into an atoll track as it decreased in luminosity
\citep{lrh09,hvfrw10}.  While it has long been known that Z sources are more
luminous than the atoll sources, this is the first time a single source has
been seen to evolve from a Z to an atoll, indicating that mass accretion rate
must drive the overall shape of the color-color diagram.  Nevertheless, it
still remains unclear as to what drives the state changes on the Z or atoll
track.  \citet{lrh09} suggest that the accretion disk evolves from a thin disk
to a slim disk, while others have previously suggested that small changes in
mass accretion rate change the source state though there is disagreement as to
which direction along the Z track mass accretion rate increases
\citep[e.g.][]{hvedm90,mmfhd07,chb06}.

In this paper, we present the results of our UV/X-ray monitoring campaign of
\cyg.  Section~\ref{sec:obs} describes the observations and data reduction.
Section~\ref{sec:analysis} we present the analysis and results, including the
surprising finding that, contrary to expectations from the reprocessing model,
the NUV and BAT X-ray flux are anticorrelated.  Finally, we discuss the
implications of our results in Section~\ref{sec:discussion}.

\section{Observations and Data Reduction}
\label{sec:obs}

\swift{} visited \cyg{} for 51 observations between 2008 June 30 and 2008
November 11, comprising observations 00090045001 through 00090045056.  XRT
observations were taken in windowed timing (WT) mode due to the high
count-rate.  The UVOT exposures were taken in two filters ($UVW1$, $UVW2$), and
BAT data was taken in standard survey mode.  Table~\ref{tab:observations}
describes the observations, exposure times, and rates for the XRT, $UVW2$, and
BAT detections described in this section.

\subsection{\label{sec:xrtreduction}XRT Data Reduction}

The XRT observations were processed using the packages and tools available in
HEASOFT version 6.6.1\footnote{See
http://heasarc.gsfc.nasa.gov/docs/software/lheasoft}.  Initial event cleaning
was performed with ``xrtpipeline'' using standard quality cuts, and event
grades 0-2 in WT mode.  For the WT mode data, source extraction was performed
with ``xselect'' in a rectangular box 20 pixels wide and 60 pixels
long. Background extraction was performed with a box 20 pixels wide and 60
pixel long far from the source region.  Several \swift{} observations contain
multiple pointings separated by more than an hour.  For these observations,
each individual pointing was processed separately, as the detector response
varies depending on the location of the source in the field of view, as well as
the fact that \cyg{} may vary significantly on these timescales.

Several individual XRT pointings have been rejected for further analysis for
two reasons.  First, we demand that each orbital good time interval (GTI)
has over 100 seconds of continuous observations.  Second, we rejected 5
pointings where \cyg{} is at the edge of the WT mode field of view and the
source extraction region is truncated.  In all there are 83 epochs used in this
analysis with exposure times ranging from 100~s to 1463~s, with a median
exposure time of $\sim500\,\mathrm{s}$.

After event selection, exposure maps were generated with ``xrtexpomap'', and
ancillary response function (arf) files with ``xrtmkarf''.  The latest response
files (v011) were used from the CALDB database.  All spectra considered in this
paper were grouped to require at least 20 counts per bin using the ftool
``grppha'' to ensure valid results using $\chi^2$ statistical analysis.  The
spectra were analyzed using XSPEC version 11.3.2ag~\citep{a96}.  Fits were
restricted to the 0.6-10~keV range due to calibration uncertainties at energies
less than 0.6~keV.  The uncertainties reported in this work are $1\sigma$
errors, obtained by allowing all fit parameters to vary simultaneously.

The observations were affected by pile-up, as the observed count rate varied
from $161 - 331\,\mathrm{ct}\,\mathrm{s}^{-1}$ (0.6-10~keV).  To correct for
pile-up, we followed the spectral fitting method described in \citet{rcccc06}
and \citet{rmst07}: using various exclusion regions centered on the source, we
refit the continuum spectrum until the fit parameters did not vary
significantly.  We found that a 10 pixel exclusion region was sufficient to
correct pile-up in the brightest epochs.  For simplicity, we use the same
exclusion region for all of the observations. We then calculate the
conversion factor to determine the non-piled-up equivalent count rate.  This is
obtained from the ratio of the arf (at 1.5~keV)
calculated with and without PSF correction.  We note that this correction
is only applied when estimating the source intensity, and is not necessary when
calculating colors, which are count rate ratios.

\LongTables
\begin{deluxetable*}{cccccccc}[!tp]
\tablewidth{0pt}
\tablecaption{\label{tab:observations}\swift{} Observations of \cyg{}}
\tabletypesize{\scriptsize}
\tablehead{
\colhead{Number\tablenotemark{a}} & \colhead{Orbit} & 
\colhead{XRT} & \colhead{XRT\tablenotemark{b}} &
\colhead{$UVW2$} & \colhead{$UVW2$} &
\colhead{BAT} & \colhead{BAT\tablenotemark{c}}\\
& & \colhead{Exp. Time (s)} & \colhead{Rate $(\mathrm{ct}\,\mathrm{s}^{-1})$} &
\colhead{Exp. Time (s)} & \colhead{Magnitude} &
\colhead{Exp. Time (s)} & \colhead{Rate $(10^{-2}\,\mathrm{ct}\,\mathrm{s}^{-1}\,\mathrm{cm^{-2}})$}
}

\startdata
001 &  1 & 1192 & $ 500.2 \pm 1.2$ & 584 & $ 15.35 \pm 0.03$ & 1200 & $  5.5 \pm 0.8$\\
002 &  1 & 1183 & $ 230.1 \pm 0.7$ & 584 & $ 14.70 \pm 0.02$ & 1211 & $  1.0 \pm 0.7$\\
003 &  1 &  337 & $ 560.5 \pm 2.4$ & 199 & $ 15.52 \pm 0.04$ &  350 & $ 10.7 \pm 1.3$\\
    &  2 &  439 & $ 353.0 \pm 1.7$ & 230 & $ 15.50 \pm 0.04$ &  460 & $  8.1 \pm 1.0$\\
004 &  1 &  912 & $ 464.0 \pm 1.3$ & 584 & $ 14.83 \pm 0.02$ &  300 & $  0.3 \pm 1.3$\\
005 &  1 & 1077 & $ 430.0 \pm 1.3$ & 525 & $ 15.39 \pm 0.03$ & 1091 & $  5.6 \pm 0.7$\\
    &  2 &  771 & $ 475.2 \pm 1.6$ & 348 & $ 15.25 \pm 0.04$ &  860 & $  2.7 \pm 0.9$\\
006 &  2 &  806 & $ 516.8 \pm 1.3$ & 436 & $ 15.70 \pm 0.03$ &  844 & $  7.0 \pm 0.8$\\
007 &  1 &  850 & $ 473.1 \pm 1.3$ & 436 & $ 15.65 \pm 0.03$ &  864 & $  7.3 \pm 0.8$\\
    &  2 &  909 & $ 341.9 \pm 1.2$ & 466 & $ 15.68 \pm 0.02$ &  924 & $  9.5 \pm 0.8$\\
    &  3 &  529 & $ 574.5 \pm 2.0$ & 289 & $ 15.62 \pm 0.03$ &  564 & $ 11.7 \pm 1.0$\\
008 &  1 &  358 & $ 425.0 \pm 2.1$ & 200 & $ 15.34 \pm 0.04$ &  372 & $  4.6 \pm 1.1$\\
    &  2 &  418 & $ 635.3 \pm 2.3$ & 230 & $ 15.68 \pm 0.04$ &  432 & $  5.6 \pm 1.1$\\
    &  3 &  718 & $ 336.6 \pm 1.4$ & 377 & $ 15.70 \pm 0.03$ &  732 & $  6.0 \pm 0.9$\\
    &  4 &  514 & $ 594.0 \pm 2.1$ & 289 & $ 15.80 \pm 0.03$ &  552 & $  8.6 \pm 1.0$\\
009 &  1 & 1068 & $ 350.0 \pm 1.0$ & 554 & $ 15.99 \pm 0.03$ & 1109 & $  8.6 \pm 0.7$\\
010 &  1 & 1079 & $ 323.7 \pm 1.0$ & 584 & $ 15.70 \pm 0.03$ & 1100 & $  9.9 \pm 0.9$\\
011 &  1 & 1136 & $ 415.1 \pm 1.1$ & 584 & $ 14.84 \pm 0.02$ & 1165 & $  3.1 \pm 0.8$\\
012 &  1 &  976 & $ 388.8 \pm 1.2$ & 495 & $ 15.61 \pm 0.02$ & 1009 & $  5.6 \pm 0.8$\\
013 &  1 &  648 & $ 423.6 \pm 1.3$ & 318 & $ 15.04 \pm 0.05$ &  663 & $  2.9 \pm 0.9$\\
    &  2 &  402 & $ 334.1 \pm 1.4$ & 199 & $ 15.00 \pm 0.05$ &  425 & $  4.4 \pm 1.1$\\
015 &  1 & 1055 & $ 325.3 \pm 0.9$ & 525 & $ 14.89 \pm 0.02$ & 1090 & $  3.7 \pm 0.7$\\
016 &  1 & 1032 & $ 414.6 \pm 1.1$ & 525 & $ 15.94 \pm 0.03$ & 1062 & $  8.0 \pm 0.8$\\
017 &  1 & 1324 & $ 473.6 \pm 1.1$ & 673 & $ 15.53 \pm 0.02$ & 1362 & $  3.7 \pm 0.7$\\
018 &  1 &  929 & $ 552.1 \pm 1.5$ & 495 & $ 15.70 \pm 0.02$ &  942 & $  5.3 \pm 0.8$\\
    &  2 &  897 & $ 319.2 \pm 1.3$ & 494 & $ 15.13 \pm 0.04$ &  903 & $  4.0 \pm 0.7$\\
019 &  1 & 1183 & $ 389.5 \pm 0.9$ & 612 & $ 15.79 \pm 0.03$ & 1200 & $  6.4 \pm 0.8$\\
020 &  1 & 1095 & $ 404.9 \pm 1.1$ & 554 & $ 15.55 \pm 0.03$ & 1129 & $  5.4 \pm 0.8$\\
022 &  1 & 1009 & $ 401.7 \pm 1.2$ & 525 & $ 14.96 \pm 0.06$ & 1024 & $  2.0 \pm 0.7$\\
    &  2 & 1108 & $ 442.5 \pm 1.2$ & 584 & $ 14.88 \pm 0.03$ & 1144 & $  2.6 \pm 0.7$\\
023 &  1 & 1020 & $ 395.2 \pm 1.0$ & 525 & $ 15.46 \pm 0.03$ & 1034 & $  6.0 \pm 0.7$\\
    &  2 &  132 & $ 445.9 \pm 2.9$ &  -- & -- &  170 & $  8.2 \pm 1.6$\\
025 &  1 &  671 & $ 453.5 \pm 1.6$ & 348 & $ 15.49 \pm 0.04$ &  684 & $  4.6 \pm 0.8$\\
026 &  1 &  946 & $ 327.2 \pm 0.9$ & 495 & $ 15.06 \pm 0.03$ &  975 & $  1.4 \pm 0.7$\\
027 &  1 & 1463 & $ 321.3 \pm 1.0$ & 731 & $ 16.03 \pm 0.03$ & 1200 & $  7.8 \pm 0.6$\\
028 &  1 & 1354 & $ 400.2 \pm 1.0$ & 671 & $ 15.64 \pm 0.02$ & 1384 & $  2.6 \pm 0.6$\\
029 &  1 &  143 & $ 299.0 \pm 2.6$ & 136 & $ 15.49 \pm 0.04$ &  157 & $  4.8 \pm 1.6$\\
    &  2 &  456 & $ 359.1 \pm 1.4$ & 258 & $ 15.31 \pm 0.04$ &  487 & $  2.7 \pm 0.9$\\
030 &  1 &  225 & $ 332.0 \pm 2.1$ & 217 & $ 16.07 \pm 0.05$ &  213 & $ 10.4 \pm 1.4$\\
    &  2 &  297 & $ 386.7 \pm 2.0$ & 286 & $ 16.02 \pm 0.04$ &  310 & $  8.7 \pm 1.2$\\
    &  5 &  298 & $ 369.5 \pm 2.0$ & 310 & $ 15.87 \pm 0.03$ &  332 & $  7.9 \pm 1.2$\\
    &  6 &  374 & $ 226.0 \pm 1.3$ &  -- & -- &  -- & --\\
    &  7 &  302 & $ 324.1 \pm 1.8$ &  -- & -- &  -- & --\\
031 &  1 & 1417 & $ 257.5 \pm 0.8$ & 702 & $ 14.97 \pm 0.04$ &  600 & $  6.0 \pm 0.9$\\
032 &  1 &  437 & $ 258.1 \pm 1.3$ & 230 & $ 15.63 \pm 0.04$ &  450 & $  7.4 \pm 1.1$\\
    &  2 &  442 & $ 240.5 \pm 1.3$ & 230 & $ 15.79 \pm 0.04$ &  474 & $  3.9 \pm 1.0$\\
033 &  1 &  647 & $ 277.2 \pm 1.0$ & 318 & $ 16.16 \pm 0.04$ &  666 & $  9.2 \pm 0.9$\\
    &  2 &  643 & $ 338.8 \pm 1.1$ & 318 & $ 16.22 \pm 0.04$ &  662 & $  8.3 \pm 0.9$\\
034 &  1 &  795 & $ 230.1 \pm 1.0$ & 407 & $ 16.24 \pm 0.04$ &  818 & $  8.4 \pm 0.8$\\
036 &  1 &  756 & $ 207.0 \pm 0.9$ & 377 & $ 16.03 \pm 0.04$ &  770 & $  9.2 \pm 0.8$\\
    &  2 &  738 & $ 190.0 \pm 0.9$ & 377 & $ 16.02 \pm 0.04$ &  770 & $  7.7 \pm 0.8$\\
037 &  1 &  798 & $ 169.7 \pm 0.7$ & 406 & $ 15.60 \pm 0.03$ &  830 & $  9.3 \pm 0.8$\\
038 &  1 & 1109 & $ 243.7 \pm 0.7$ & 553 & $ 15.83 \pm 0.03$ & 1150 & $  6.1 \pm 0.7$\\
039 &  1 & 1052 & $ 208.0 \pm 0.7$ & 524 & $ 16.11 \pm 0.03$ & 1086 & $  5.5 \pm 0.7$\\
041 &  1 &  142 & $ 210.2 \pm 2.0$ &  -- & -- &  156 & $  2.9 \pm 1.6$\\
    &  2 &  142 & $ 215.9 \pm 2.0$ &  -- & -- &  156 & $  9.8 \pm 1.7$\\
    &  3 &  142 & $ 184.4 \pm 2.2$ &  -- & -- &  153 & $  4.7 \pm 1.7$\\
    &  4 &  142 & $ 210.1 \pm 2.0$ &  -- & -- &  156 & $  7.7 \pm 1.7$\\
    &  5 &  254 & $ 204.7 \pm 1.5$ & 111 & $ 16.25 \pm 0.07$ &  268 & $  7.8 \pm 1.2$\\
    &  6 &  262 & $ 179.5 \pm 1.4$ & 140 & $ 15.45 \pm 0.04$ &  276 & $  4.3 \pm 1.2$\\
    &  7 &  143 & $ 211.6 \pm 2.0$ &  -- & -- &  156 & $  3.6 \pm 1.6$\\
    &  8 &  142 & $ 190.5 \pm 1.9$ &  -- & -- &  156 & $  8.5 \pm 1.8$\\
    &  9 &  143 & $ 164.9 \pm 1.9$ &  -- & -- &  157 & $  7.9 \pm 1.7$\\
    & 10 &  131 & $ 251.4 \pm 2.5$ &  -- & -- &  156 & $  5.1 \pm 1.6$\\
042 &  1 &  409 & $ 243.8 \pm 1.3$ & 199 & $ 15.74 \pm 0.04$ &  422 & $  2.8 \pm 1.0$\\
    &  2 &  471 & $ 250.3 \pm 1.3$ & 230 & $ 15.69 \pm 0.03$ &  485 & $  6.1 \pm 1.0$\\
    &  4 &  471 & $ 260.2 \pm 1.3$ & 230 & $ 16.15 \pm 0.05$ &  485 & $  5.1 \pm 0.9$\\
044 &  2 &  211 & $ 211.3 \pm 1.6$ & 220 & $ 16.28 \pm 0.05$ &  241 & $ 11.0 \pm 1.4$\\
045 &  2 &  336 & $ 308.7 \pm 1.6$ & 171 & $ 15.71 \pm 0.05$ &  349 & $  7.2 \pm 1.1$\\
    &  3 &  249 & $ 329.3 \pm 2.0$ & 141 & $ 15.70 \pm 0.05$ &  265 & $  5.0 \pm 1.3$\\
046 &  1 &  498 & $ 380.7 \pm 1.5$ & 258 & $ 15.86 \pm 0.04$ &  511 & $  6.5 \pm 0.9$\\
    &  2 &  417 & $ 355.8 \pm 1.6$ & 229 & $ 15.25 \pm 0.04$ &  451 & $  2.6 \pm 1.0$\\
047 &  1 &  378 & $ 416.9 \pm 1.8$ & 199 & $ 15.62 \pm 0.04$ &  -- & --\\
    &  2 &  784 & $ 432.0 \pm 1.3$ & 406 & $ 15.19 \pm 0.03$ &  811 & $  2.8 \pm 0.8$\\
048 &  1 &  465 & $ 574.9 \pm 2.1$ & 230 & $ 15.80 \pm 0.04$ &  479 & $ 10.5 \pm 1.0$\\
049 &  1 &  465 & $ 523.2 \pm 1.9$ & 230 & $ 15.65 \pm 0.04$ &  478 & $  8.2 \pm 1.0$\\
    &  2 &  465 & $ 525.6 \pm 1.9$ &  -- & -- &  478 & $ 10.5 \pm 1.0$\\
    &  3 &  445 & $ 426.2 \pm 1.8$ & 229 & $ 15.87 \pm 0.04$ &  478 & $ 10.4 \pm 1.0$\\
053 &  1 &  493 & $ 515.4 \pm 1.7$ & 259 & $ 16.05 \pm 0.05$ &  505 & $ 10.0 \pm 0.9$\\
    &  2 &  478 & $ 520.6 \pm 1.8$ & 259 & $ 16.08 \pm 0.06$ &  505 & $  9.1 \pm 1.0$\\
055 &  1 &  947 & $ 484.6 \pm 1.3$ &  -- & -- &  -- & --\\
056 &  1 &  548 & $ 461.1 \pm 1.6$ &  -- & -- &  -- & --\\
    &  2 &  408 & $ 425.2 \pm 1.8$ &  -- & -- &  -- & --\\
\enddata
\tablenotetext{a}{The full observation number is given by prepending 00090045.}
\tablenotetext{b}{In the 0.6-10 keV range.}
\tablenotetext{c}{In the 14-24 keV range.}
\end{deluxetable*}

\subsection{\label{sec:uvreduction}UVOT Data Reduction}

The UVOT analysis we performed is similar to that of \citet{rmst07}.  The UVOT
images were initially processed at HEASARC using the standard \swift{}
``uvotpipeline'' procedure, with standard event cleaning.  The initial
astrometric solution of UVOT images is typically offset by up to 5-10\arcsec.
We corrected for this offset by matching stars with the USNO B1.0 catalog,
improving the aspect solution to better than 1\arcsec.  Our procedure is
similar to the ftool ``uvotskycorr''.  Due to the relatively rapid variation of
\cyg{}, observations with more than one exposure were analyzed independently.

The UVOT images are not very crowded, in spite of the relatively low Galactic
latitude ($-11.3^\circ$), although a bright star just outside the field of view
created a noticeable blaze in the images.  Initial photometry was performed
using ``uvotdetect'' with the calibration option set to perform coincidence
loss correction and calibration to standard UVOT photometry~\citep{pbplh08}.
We then perform relative photometry using 36 template stars that are well
measured and brighter than 17.5 mag in the $UVW2$ filter.  The rms error for
the relative photometry correction is typically $\sim5\%$, and we confirm that
the corrected light curves of the comparison stars are stable within $\sim5\%$.
At this step we remove bad observations where fewer than 3 template stars are
detected.  These are typically observations with exposure times less than
$\sim100\,\mathrm{s}$.  After rejecting 10 bad observations, we observed \cyg{}
for a total of 68 epochs in the $UVW2$ filter.  There were fewer (46) good
observations with the $UVM2$ filter.  Therefore, we concentrate our further
analysis on the broad $UVW2$ filter, although we have confirmed that our
results are identical for $UVM2$, which has a substantial wavelength overlap
with $UVW2$.

\subsection{\label{sec:batreduction}BAT Data Reduction}

The \swift/BAT analysis was performed with the ftool ``batsurvey,'' which takes
as input the BAT detector plane histograms (DPHs) assembled in survey mode, and
outputs background subtracted flux values.  We ran ``batsurvey'' in SNAPSHOT
mode to combine all DPHs from each individual orbit, and set the binning to the
standard 8 energy channels
(14-20,20-24,24-35,35-50,50-75,75-100,100-150,150-195 keV).  An input source
catalog containing the location of \cyg{} was created to ensure that the flux
would be estimated even when it was not detected at a $3\sigma$ level.  Similar
to the analysis for the XRT and UVOT data, we demanded a minimum exposure time
of $100\,\mathrm{s}$, which yields 77 epochs. The spectrum of \cyg{} is very
soft in the BAT band, and it is generally not detected in the hardest X-ray
channels.  By summing over energy channels, we were able to determine that the
BAT flux in the 14-24 keV range provides the maximum signal-to-noise for the
majority of the BAT measurements.  Finally, we convert the ``batsurvey'' output
rates to $\mathrm{ct}\,\mathrm{cm^{-2}}\,\mathrm{s}^{-1}$.

\section{Analysis \& Results}
\label{sec:analysis}

\subsection{Color-Color Diagram}

The most useful method of displaying the variability of Z-sources such as
\cyg{} is with a color-color diagram.  The bandpass of \swift/XRT is narrower
and softer than other X-ray detectors such as \rxte/PCA, so we are unable
to use the color definitions commonly applied to \rxte{} data.  We follow
the color definitions of \citet{shjny09}, who defined colors appropriate to
\emph{Chandra}/HETGS, which has a similar soft bandpass as \swift/XRT.  The
three color bands we define are 0.6-2.5 keV (soft), 2.5-4.5 keV (middle), and
4.5-10 keV (hard), such that the soft (hard) color is defined as the ratio of
the count rate in the middle-to-soft (hard-to-middle) bands.

\begin{figure}
\centering
\scalebox{0.85}{\rotatebox{270}{\plotone{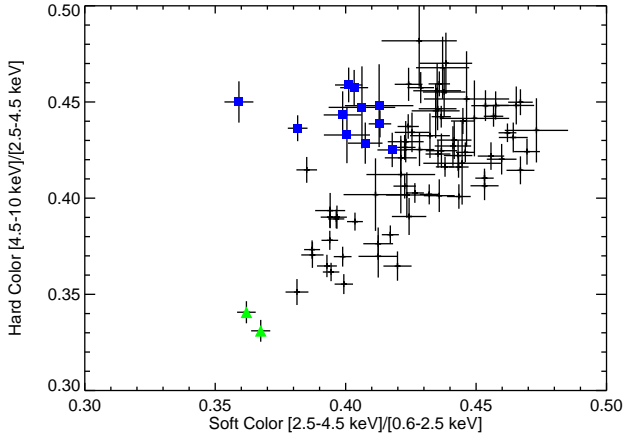}}}
\caption{\label{fig:cc}Color-color plot for 83 epochs of \cyg{} observed with
  XRT.  Most of the observations are along the NB and vertex of
  the NB with the HB.  The HB observations are marked with
  squares, and the FB observations are marked with triangles.}
\end{figure}

\begin{figure}
\centering
\scalebox{0.85}{\rotatebox{270}{\plotone{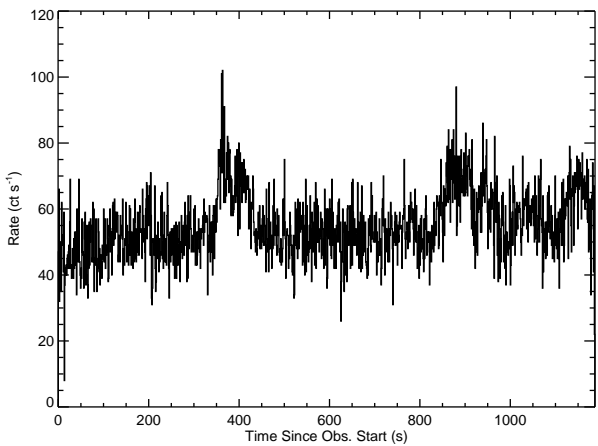}}}
\caption{\label{fig:fblc}Light curve of \cyg{} during observation 002 in the
  flaring branch.  The low count rate and high variability is typical of the
  FB, and is not observed in the NB and HB.}
\end{figure}

The resulting color-color diagram is shown in Figure~\ref{fig:cc}.  Most of our
observations are along the NB or the vertex of the NB and the HB.  For visual
reference in the succeeding figures, we have marked the observations along the
HB based on soft color (sc) and hard color (hc), using an arbitrary definition of $\mathrm{sc}<0.42$ and
$0.42<\mathrm{hc}<0.46$.  The two observations that have the softest colors are
clearly on the FB.  In addition to being outliers in the
color-color plot, the light curves of these observations show strong
variability characteristic of the FB.  A sample light curve from
observation 002 in the FB is shown in Figure~\ref{fig:fblc}.  Thus,
over the course of 83 epochs we have snapshots scanning the entire Z-track of
\cyg{}.

\subsection{Simultaneous Soft X-ray and NUV Observations}

\begin{figure}
\centering
\scalebox{1.1}{\plotone{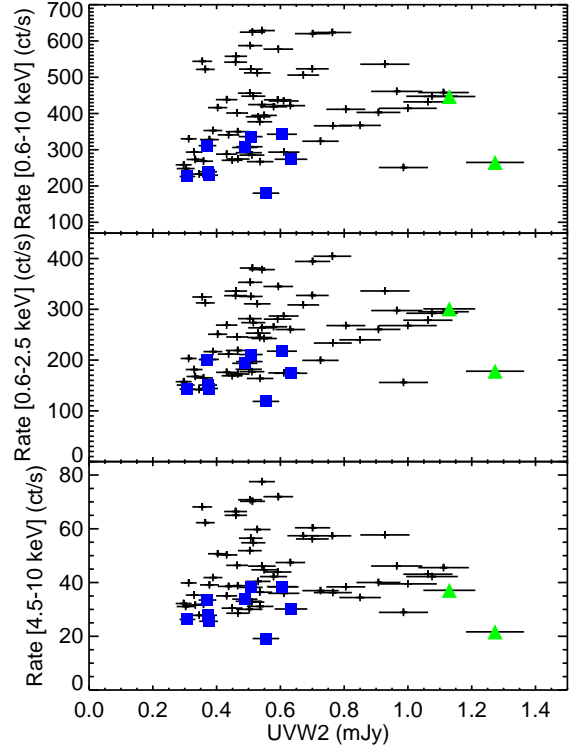}}
\caption{\label{fig:rateuv}XRT rate in three different bands vs. $UVW2$ flux
  density.  The top, middle, and bottom panels show the soft, medium, and hard
  XRT flux respectively.  There is no clear correlation of XRT rate with NUV
  flux density, although the outline of the Z-track can be inferred, especially
  in the bottom panel.}
\end{figure}

The multi-wavelength \swift{} observatory allows us to easily obtain
simultaneous observations of \cyg{} in both X-ray and NUV wavelengths.  If the
NUV emission is reprocessed X-ray emission, we may expect the NUV flux to track
the X-ray flux, as described in \S~\ref{sec:intro}. Figure~\ref{fig:rateuv}
shows the XRT rate in the ``soft'', ``medium'', and ``hard XRT'' energy bands
as a function of $UVW2$ flux density.  We do not observe an obvious
correlation between the NUV and the X-ray flux for energies less than
$10\,\mathrm{keV}$.  However, we do notice the Z-track apparent in the figure,
especially for the bottom panel which shows the hard X-ray band vs. $UVW2$.
Particularly notable is the fact that the FB observations correlate with the
brightest NUV observations, and the HB observations correlate with some of the
dimmest NUV observations.  A similar trend was noted with 7 simultaneous
observations using \emph{Ginga} and \emph{IUE}~\citep{vrgvh90}.  Those authors
concluded that the NUV flux is brightest on the FB and dimmest on the HB.

\begin{figure}
\centering
\scalebox{1.1}{\plotone{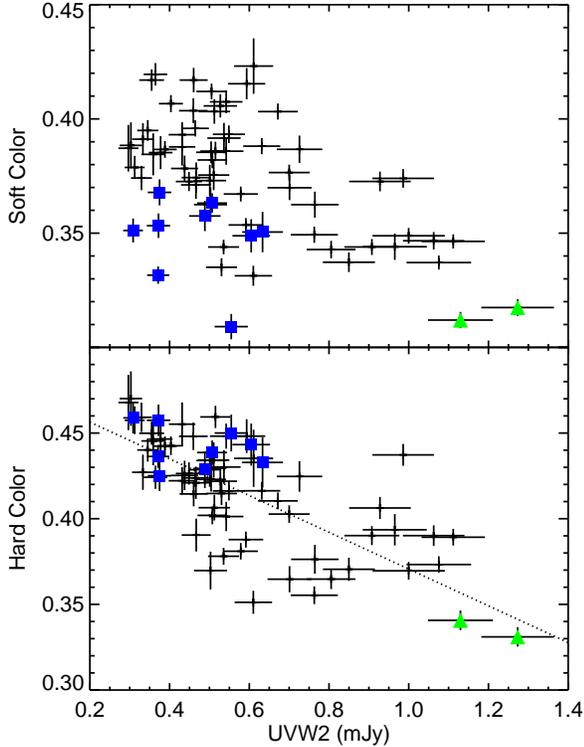}}
\caption{\label{fig:coloruv}Soft X-ray color (top) and Hard X-ray color
  (bottom) vs. $UVW2$ flux density.  There is no obvious correlation between
  the soft color and the NUV, but there is a strong anticorrelation between
  the hard color and the NUV. The best-fit line, shown with a dotted line, has
  the functional form $\mathrm{hc} = 0.429\pm0.003 -
  0.11\pm0.01(f_{\nu,UVW2}-0.5)$.  The Spearman correlation coefficient,
  $r=-0.69$, shows a significant anticorrelation.}
\end{figure}

We next investigate if there is any correlation between X-ray color and NUV
flux.  Figure~\ref{fig:coloruv} shows the hard and soft X-ray color vs. the
$UVW2$ flux density.  Although there is no strong correlation between the soft
X-ray color and the NUV, there is a strong anticorrelation between the hard
X-ray color and the NUV.  That is, as the hard X-ray color decreases, the NUV
flux increases and vice-versa.  This anticorrelation drives the apparent
correlation between NUV flux and Z-track location observed by \citet{vrgvh90}:
the FB has the softest X-ray color and highest NUV flux, and the HB has a
harder X-ray color and lower NUV flux.  However, our complete coverage of the
Z-track of \cyg{} reveals that it is the hard color, not the position along the
Z-track, that is anticorrelated with the NUV flux.  Otherwise, we would observe
the NUV flux to vary along the HB, rather than to vary with hardness.

We have fit the hard color--$UVW2$ relation with the functional form
$\mathrm{hc} = \alpha + \beta(f_{\nu,UVW2}-0.5)$ using the {\tt linmix} linear
regression package~\citep{k07}.  We find that $\alpha = 0.429\pm0.003$ and
$\beta=0.11\pm0.01$, with the best-fit slope less than 0 at the $\sim10\sigma$
level.  We further confirm the significant anticorrelation by calculating the
Spearman correlation coefficient, such that $r=-0.69$.  Finally, looking at the
bottom panel of Figure~\ref{fig:rateuv}, we can see a hint of this
anticorrelation, as the NUV flux density is noticeably anticorrelated with
the 4.5-10 keV flux when \cyg{} is observed in the NB and FB.

\subsection{Simultaneous Soft and Hard X-ray and NUV Observations}

\begin{figure}
\centering
\scalebox{1.1}{\plotone{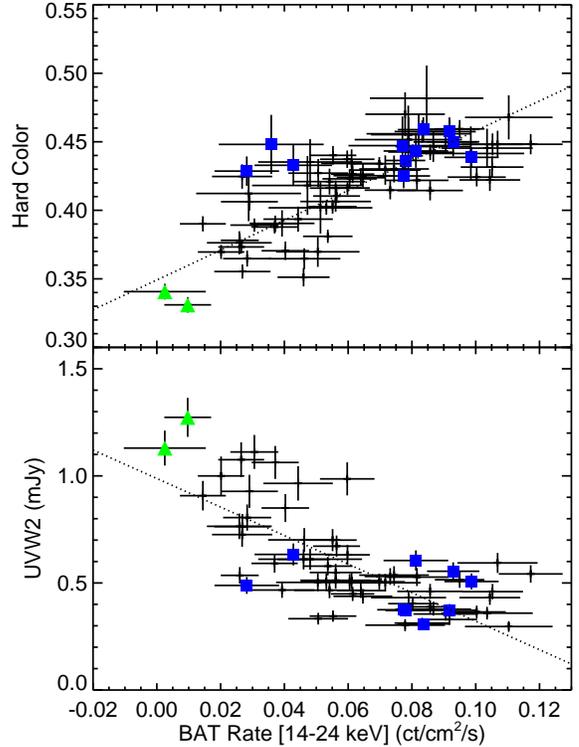}}
\caption{\label{fig:colorbat}Hard X-ray color (top) and $UVW2$ flux density
  (bottom) vs. hard X-ray rate, as measured by BAT in the 14-24 keV range.  The
  hard X-ray color and BAT rate are strongly correlated, and the $UVW2$ flux
  density and BAT rate are anticorrelated.}
\end{figure}

We now compare the XRT and NUV data with the hard X-ray observations of \cyg{}
obtained with BAT.  As discussed in \S~\ref{sec:batreduction}, \cyg{}
has a relatively soft spectrum and is only significantly observed in first two
channels, corresponding to the 14-24 keV energy band, which we refer to as
``BAT X-rays''.  The top panel of Figure~\ref{fig:colorbat} shows the hard
X-ray color (as measured by the XRT) against the BAT rate.  These two
quantities are very strongly correlated, with a Spearman coefficient of
$r=0.84$.  Fitting to the functional form $\mathrm{hc} = \alpha + \beta
(r_\mathrm{BAT}-0.05)$, where $r_\mathrm{BAT}$ is the BAT rate measured in
$\mathrm{ct}\,\mathrm{cm}^{-2}\,\mathrm{s}^{-1}$, we find $\alpha =
0.407\pm0.003$ and $\beta = 1.14\pm0.10$.  This strong correlation shows that
the hard X-ray rate is dominated by the spectral index in the 2.5-10 keV band
-- a steeper spectral index yields a softer color and a lower hard X-ray rate.
We have also confirmed that the BAT rate also depends on the XRT rate in the
4.5-10 keV band, although the hard color is the dominant driver.

The bottom panel of Figure~\ref{fig:colorbat} shows the $UVW2$ flux density
against the BAT rate.  As with the $UVW2$ flux density and the hard X-ray
color, these two quantities are strongly anticorrelated, with a Spearman
coefficient of $r=-0.68$.  The best-fit line to the functional form
$f_{\nu,UVW2} = \alpha + \beta(r_\mathrm{BAT}-0.05)$ yields $\alpha =
0.66\pm0.02$ and $\beta = -0.67\pm0.9$.  As with the anticorrelation between
the hard color and the NUV flux, there is no indication that the NUV flux
varies along the HB.  It is difficult to reconcile the observed anticorrelation
with a simple model in which the NUV emission is hard X-ray emission
reprocessed by the accretion disk.  We discuss these implications in
Section~\ref{sec:discussion}.

\subsection{Spectral evolution}
\label{sec:spectral}

As described in Section~\ref{sec:intro}, the spectral evolution of sources
along their Z track is of great interest, and it is still not entirely clear as
to the mechanism that drives these spectral changes.  \cyg{} has been the focus
of several previous investigations into Z track spectral evolution
\citep[e.g.,][]{hvedm90,vrgvh90,dzs02,psk02,dfbfk02,vrbms03}.  These previous
observations have been performed with a variety of X-ray missions including
\emph{Ginga}, \rxte, and \emph{BeppoSAX}.  These three missions, in particular
had/have quite broad energy coverage from a few keV to greater than 20 keV,
allowing for a good characterization of the continuum, though lacking the
spectral resolution to study any line features in detail.  These previous
studies find that the spectrum of \cyg{} can be fit by a variety of different
models (as is usual for Z sources).  For instance, \citet{hvedm90} fit the
\emph{Eastern} model to \emph{Ginga} data and find temperatures for the
single-temperature blackbody component of 2.2-2.7 keV, and temperatures for the
disk blackbody of 1.5-1.9 keV.  \citet{dfbfk02} fit a disk blackbody plus a
Comptonized component (XSPEC model ``comptt'') to \emph{BeppoSAX} observations
and found disk temperatures from 0.8-1.7 keV and plasma temperatures upwards of
3~keV.

The observations of \citet{hvedm90} cover all three spectral states (HB, NB,
and FB).  They note that the largest change in spectral shape is seen as the
source goes from the NB to the FB, where the disk temperature and luminosity
increase significantly when modeled by the \emph{Eastern} model.  The
observations of \citet{dfbfk02} cover mostly the HB and the NB. These authors
suggest that the inner rim of the accretion disk approaches the neutron star
surface as the source moves from the HB to the NB.



In an attempt to further understand the continuum spectral evolution of \cyg{},
we examined the \swift/XRT spectra from our monitoring campaign.  In the 0.6-10
keV band, a good fit to the XRT data can be achieved using a two thermal
component model (disk blackbody and a blackbody).  For this energy range, no
additional power-law component is required.  For the photoelectric absorption,
we fix the column density to $N_H = 1.9\times10^{21}$ cm$^{-2}$.  This is
consistent with values determined from both HI observations\footnote{using the
HEASARC $N_H$ tool: http://heasarc.gsfc.nasa.gov/cgi-bin/Tools/w3nh/w3nh.pl},
and from previous fits to the X-ray spectra \citep[e.g.][]{dfbfk02}.  With this
model, we consistently find temperatures for the disk blackbody and blackbody
components of around $\sim0.5$ and $\sim1.0-1.5$ keV respectively.  Note that
these temperatures are significantly lower than seen by the previous spectral
studies of \cyg{} discussed above.

In order to address this issue, we searched for any \rxte{} observations that
were simultaneous with any of our XRT observations.  We found that our
\swift{} observation 002, orbit 1 was overlapping with \rxte{} observation
93443-01-01-15.  The XRT observation was performed on 2008-07-02 from
23:26 to 23:46, whereas the \rxte{} observation ran from 23:00 to 23:38 on the
same date.  We extracted the \rxte/PCA spectrum from PCU 2 only (the most
reliable PCU), using the standard goodtime filtering and deadtime corrections.
We use the Standard 2 mode data, applying a systematic error of 0.6\% to each
channel of the spectrum \citep[we follow the same method as][for the \rxte{}
data reduction]{cmhvl09}.

First, we have fit the two thermal component model to the XRT data in the
0.6-10 keV energy range.  We find an inner disk temperature of
$kT_{\mathrm{in}} = 0.35\pm0.01$~keV and blackbody temperature of
$kT_{\mathrm{bb}} = 1.00\pm0.01$~keV (all errors at the $1\sigma$ level).
However, when fitting the same model\footnote{We add a Gaussian to model the Fe
  K$\alpha$ line not detectable in the XRT spectrum.} to the \rxte spectrum in
the 3-23 keV band, we find $kT_{\mathrm{in}}=1.28\pm0.01$~keV and
$kT_{\mathrm{bb}} = 2.08\pm0.03$~keV, similar to what has been observed in previous
studies of \cyg{}.  The higher energy coverage of \rxte/PCA is much better
suited to constrain these thermal components which have a peak energy of
$\sim4\,\mathrm{keV}$ and $\sim6\,\mathrm{keV}$ respectively.

\begin{figure}
\centering
\scalebox{0.7}{\rotatebox{270}{\plotone{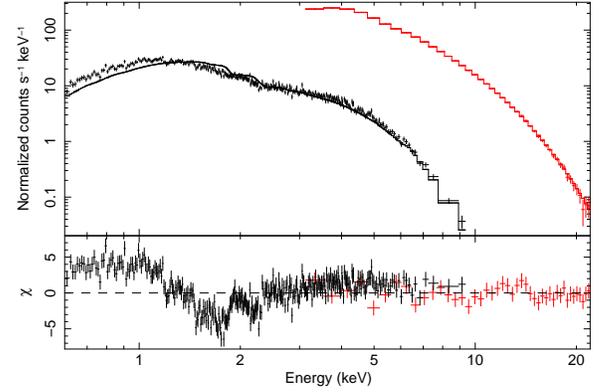}}}
\caption{Joint spectral fit between \swift/XRT (black) and \rxte/PCA
  (red). Although the spectra are in good agreement in the overlap region above
  3 keV, there are significant residuals in the XRT spectrum below 3 keV.}
\label{fig:jointspec}
\end{figure}

To test the source of this discrepancy, we have fit the XRT and \rxte{} spectra
jointly in the 0.6-23 keV range.  We tie all model parameters in the absorbed
two thermal component model between the two data sets.  We also add a constant
factor to allow for any absolute flux calibration mismatch.  The resulting
spectral fit is shown in Figure~\ref{fig:jointspec}, and is clearly a bad fit,
with a reduced $\chi^2$ of 3.5.  The best-fitting model returns temperatures
consistent with the parameters we find when fitting the \emph{RXTE} data alone,
and the \emph{spectra match nicely where they overlap in the 3-10 keV range}.
However, there are large residuals present below 3 keV.  After trying a wide
range of other models (including combinations of disk blackbody, blackbody,
power-law and Comptonization) we were unable to find a good fit.  Allow $N_H$
to float as a free parameter did improve the fit, but not to an acceptable
level, and also returned a very low value ($\sim
0.9\times10^{21}\,\mathrm{cm}^{-2}$).

\begin{figure}
\centering
\scalebox{0.8}{\rotatebox{270}{\plotone{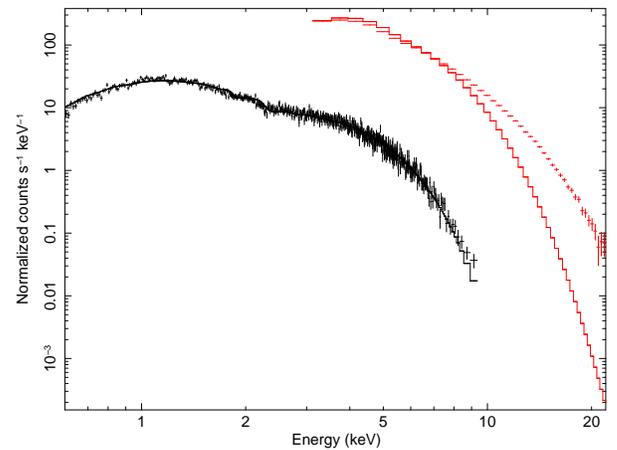}}}
\caption{An extrapolation of the best-fit to the \swift/XRT spectrum (black)
  compared to the \rxte/PCA spectrum (red).  The temperature of the blackbody
  component in the XRT spectrum is significantly underestimated, and therefore
  the extrapolation of the XRT model significantly underpredicts the hard 10-20
  keV X-ray spectrum.}
\label{fig:xtrapol}
\end{figure}

By fitting the XRT data alone we are able to minimize the residuals below 3
keV.  However, an extrapolation of the XRT spectrum to higher energies
significantly underpredicts the hard X-rays, as shown in
Figure~\ref{fig:xtrapol}.  In general, the 10-20 keV spectrum of NSXRBs can be
well described by a 2-3 keV blackbody~\citep{cmbbb09}.  The XRT fit is
dominated by the spectral shape of the soft X-rays, where the response function
peaks.  Therefore, the fit to the XRT spectrum alone, in which the temperature
of the blackbody component is grossly underestimated, cannot be properly
extrapolated to the 10-20 keV range.

We have checked that the difference between the XRT and \rxte{} spectra does
not appear to be a cross-calibration issue between the two instruments.  As a
confirmation, we analyzed several near-simultaneous XTE and \rxte{}
observations of the black hole candidate LMXB \xte{} during
its 2006 outburst \citep[see][for details on all \swift{}
observations]{rmst07}.  The \swift/XRT spectra of this object can be fit by a
simple disk blackbody plus a power-law over a wide range in luminosity
\citep{rmst07}.  When looking at near-simultaneous \rxte{} observations, we
find that apart from a slight offset in absolute flux calibration, the spectra
have the same shape -- the power-law index and disk temperature recovered from
fitting them separately and jointly are consistent.  In the case of the neutron
star binary \cyg, the spectral decomposition is much more complicated than that
for black hole binaries.  Therefore, it seems likely that the difficulties in
using the XRT spectrum alone arises due to the multi-component spectrum of
\cyg, rather than cross-calibration issues.

A Gaussian feature at around 1 keV has been reported by previous studies of
\cyg{} \citep[e.g.][]{dfbfk02}, and including a Gaussian at around 1 keV in the
model does improve the fit.  The origin of such a spectral feature is unclear,
but recently \citet{shjny09} very briefly note that there is a complex line
blend around 1 keV in their {\it Chandra} gratings spectra, the study of which
will be the focus of their future work.  However, even with the inclusion of a
Gaussian in our fit, from fitting the XRT data alone, we still recover low
temperatures.  Therefore, we chose not to investigate the spectral
evolution of \cyg{} along the Z track with the current \swift{} data set.


\subsection{Periodicity}
\label{sec:periodicity}

Using long baseline observations of \cyg{} with \emph{RXTE}, \citet{wks96}
detected a $\sim78$ day period in the 1.5-12 keV light curve.  At different
times, a similar period of $\sim70-80$ days has been detected in \emph{Vela 5B},
\emph{Ariel 5}, and \emph{RXTE}-ASM data in addition to a shorter $\sim40$ day
period~\citep{pkm00,cccl03}.  The present set of \swift{} observations,
spanning $\sim130\,\mathrm{days}$, is sufficiently long to confirm the
ephemeris presented in \citet{wks96}.  Unfortunately, the \swift{} coverage is
not long enough to independently measure the ephemeris.

\begin{figure}
\centering
\scalebox{1.1}{\plotone{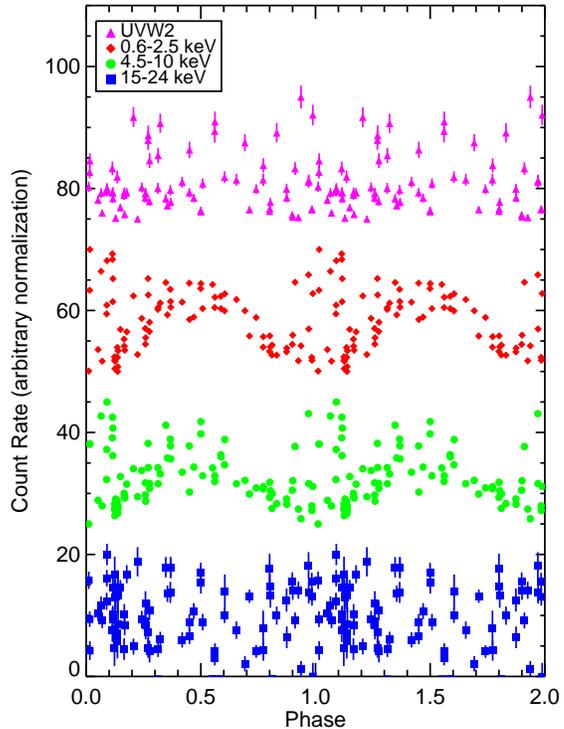}}
\caption{\label{fig:phased}Phased light curve using ephemeris from
  \citet{wks96}. The count rate in each energy band has been scaled to the same
  range, and offset for clarity.  The fractional amplitude for each energy
  range is 55\% (0.6-2.5 keV); 60\% (4.5-10 keV); 96\% (14-24 keV); and 65\%
  (\emph{UVW2}).  The period is most prominent in the soft 0.6-2.5 keV X-rays
  (red diamonds), and is not visible in the 14-24 keV BAT X-rays (blue squares)
  or the \emph{UVW2} observations (magenta triangles).}
\end{figure}

The best-fit ephemeris from \citet{wks96} is:
$$
\mathrm{JD}2442209.0 \pm 4.7 + N(77.79\pm0.08).
$$ Figure~\ref{fig:phased} shows the phased light curve using this ephemeris
for \cyg{} for two XRT energy ranges (0.6-2.5 keV; 4.5-10 keV), BAT X-rays
(14-24 keV), and \emph{UVW2}.  The count rate in each energy band has been
scaled to the same amplitude, and offset for clarity.  The true fractional
amplitude for each energy range is 55\% (0.6-2.5 keV); 60\% (4.5-10 keV); 96\%
(14-24 keV); and 65\% (\emph{UVW2}).

The phased light curve for the soft X-rays (0.6-2.5 keV; red diamonds) shows
that the period and zero-point for the ephemeris, which was measured over 13
years ago, are well-matched to the current data set.  Although there is
significant variability in addition to the underlying periodicity, we observe a
maximum at a phase of 0.5 and a minimum at a phase of 1.0.  The minimum is
consistent with previous observations, although we do not observe a less
prominent secondary minimum at a phase of 0.5~\citep{wks96}. The periodicity is
most prominent in the soft (0.6-2.5 keV) X-rays, and is less prominent in the
harder (4.5-10 keV; green circles) X-rays, and is not apparent in either the
BAT X-rays (14-24 keV; blue squares) or \emph{UVW2} (magenta triangles) data.

While the $\sim78$ day period is readily apparent in the present soft X-ray
data, we have not been able to confirm the $\sim40$ day period observed at
different times by other satellites~\citep[e.g.][]{pkm00, cccl03}.
\citet{cccl03} have suggested that the $\sim40$ day and $\sim80$ day periods
are harmonics, with the $\sim40$ day period as more stable.  By
studying \cyg over a longer time baseline than \citet{wks96}, they find that
the dominant period changes, and suggest that the long-term ephemeris is not
the reliable clock of \citet{wks96}.  Our \swift data provides more evidence
that the dominant period changes over time.  In addition, as our present data
is consistent with the ephemeris of \citet{wks96}, suggesting some long-term
stability of this underlying mode.

\citet{wks96} attribute the observed superorbital periodicity to precession of
a tilted accretion disk observed from a relatively high inclination angle,
while \citet{cccl03} attribute it to a precessing warped accretion disk.  Our
present observations are consistent with either interpretation: the accretion
disk should dominate the spectrum in the soft X-rays, while the blackbody
emission from the boundary layer dominates the BAT X-rays.  Thus, we do not
expect to observe this periodicity in the 14-24 keV range, which is dominated
by blackbody emission~\citep[e.g.,][]{cmbbb09}.

\section{Discussion}
\label{sec:discussion}

\citet{hvedm90} studied \cyg{} with a joint X-ray/NUV monitoring campaign, and
concluded that the NUV emission is an indirect but superior measure of
bolometric flux.  They argue that the accretion disk is a better bolometer than
our X-ray detectors because it is insensitive to bandpass effects.  Assuming
the NUV emission is dominated by reprocessed hard X-ray emission, it is then a
good proxy for the total X-ray flux.  With seven simultaneous X-ray/NUV
observations, there was a hint that the UV continuum flux increases along the Z
track.  They therefore conclude that $\dot m$ increases along the Z track from
the FB through the NB to the HB.

However, our current observations tend to contradict the interpretation of
\citet{hvedm90}.  First, we observe a strong anticorrelation between the NUV
flux and 14-24 keV BAT X-ray flux, which is difficult to explain in the context
of a simple reprocessing model.  Second, there is no evidence in the X-rays
that the bolometric luminosity is increasing as the BAT X-ray luminosity is
decreasing.  We observe no correlation between the XRT (0.6-10 keV) flux and
the BAT flux, even after correcting for the observational phase. Third, with 68
epochs of simultaneous NUV and X-ray observations of \cyg{}, we can demonstrate
that the NUV does not vary monotonically along the Z-track, but rather varies
inversely with the BAT X-ray emission and hard color.

The observed anticorrelation is in stark contrast to the strong correlation
observed between the NUV flux and the 2-10 keV X-ray emission observed for the
BHXRB \xte{}~\citep{rmst07}.  The most obvious difference between these two
systems is the existence of the neutron star surface and boundary layer in
\cyg{} which is not present in \xte{}.  This issue is addressed in greater
detail below. Another difference is the observed energy range.  We have newly
analyzed the 14-24 keV BAT observations of \xte{}, which was detected at the
$\sim2-3\sigma$ level during the first $\sim10$ epochs.  We have confirmed that
the 2-10 keV flux detected with the XRT is a good proxy for the BAT X-ray
emission.  The simple spectral decomposition of the BHXRB makes this possible:
while the softer X-ray emission is dominated by flux from the accretion disk,
this provides the seed photons for the hard Comptonized component. It is this
hard component that is reprocessed by the disk into the NUV emission that we
observe.


As we have discussed in Section~\ref{sec:spectral}, the spectral decomposition
of a NSXRB like \cyg{} is much more complicated.  The hard X-ray emission above
10 keV is well described by a blackbody potentially from the NS boundary
layer~\citep[e.g.,][]{cmbbb09,rg06}.  The emission from the boundary layer is not
directly correlated with the softer emission from the accretion disk, and
therefore the 2-10 keV flux is not a good proxy for the hard 14-24 keV BAT flux
in this case.  It is this hard X-ray emission that fluoresces the iron
line~\citep{cmbbb09} and should be reprocessed into NUV emission.  The more
complicated spectral decomposition of the NSXRB explains why we see a different
relationship between XRT-NUV and BAT-NUV.  However, it does not explain the
anticorrelation between the BAT and NUV flux.

A possible explanation of the large difference between the NUV$-$hard
X-ray correlation in \cyg{} and \xte{} is geometric.  Although we do not have
any constraints on the inclination of \xte{}, there is a large amount of
evidence that we are observing \cyg{} at a high inclination angle.  First, the
short duration dips in the X-ray light curve imply a high inclination
angle~\citep{vskk88}.  In addition, the periodicity in the optical light curve
has been fit with an ellipsoidal model which constrains the inclination angle
$i\sim65^\circ$ \citep{ok99}.  Finally, the long period in the soft X-ray light curve can be
attributed to the precession of a tilted accretion disk only if we are
observing the source relatively edge-on~\citep{wks96,cccl03}.

Modeling the accretion disk atmosphere and corona of a NSXRB, \citet{jrl02}
have shown that the atmosphere and corona expands as the disk is radiatively
heated.  As the corona expands, more of the reprocessed NUV flux will be
scattered out of the line-of-sight of an observer at a high inclination
angle.  The effect of the scattering may be larger than the increase in
reprocessed emission as the disk is heated by the hard X-rays.  Thus, we can
observe an anticorrelation between NUV flux and BAT X-rays in one source and a
correlation in another depending on inclination angle.

We can compare our present observations to those of the BHXRB GX339$-$4, which
was observed simultaneously in the optical and X-rays~\citep{mrpic83}.  They
observe an anti-correlation between the softer X-rays (1-13 keV) and the
optical, which is qualitatively different than observed for the NUV/X-ray
observations of both \cyg~(uncorrelated) and \xte~(correlated).  However, the
hard X-rays (13-20 keV) were observed to have a significantly different
behavior, in that they were slightly \emph{correlated} with the optical
emission.  As GX339$-$4 is observed at a relatively low inclination
angle~\citep{cshc02}, these observations are generally consistent with our
geometric interpretation.

There are some remaining issues with this simple picture.  The dips in the
X-ray light curves occur primarily in the FB when the NUV emission is
brightest.  This may be explained by a cooling and condensing corona which
begins to clump as it cools.  The dips would then be a signature of this cool,
clumpy corona.  When the corona is heated and starts to puff up along the NB
and HB, we scatter more NUV emission but the soft X-rays are no longer
absorbed.  Another issue is that the boundary layer could potentially be
blocked by the high inclination. 

Recently, \citet{shjny09} obtained long \emph{Chandra} gratings spectra of
\cyg{} throughout the Z track.  They find a variety of broad emission lines in
the spectra, and discovered that the line fluxes increase along the Z track
(lowest on the HB, highest on the FB).  They suggest that this implies the
average heating luminosity of the accretion disk therefore increases from HB to
NB to FB.  However, if the line emission is not fluorescent, but rather
recombination emission, we may only be able to see the lines when the heating
and ionization level of the gas is reduced.  Further work will have to be done
to fully develop this model.

Another possible explanation of the NUV$-$hard X-ray anti-correlation is given
by the spectral modeling of \citet{rg06}, applied to several Z-track NSXRBs.
Using frequency resolved spectroscopy to decompose the spectral
components~\citep{grm03}, they were able to separate the component
varying on s-ms timescales, which has a harder spectrum and is consistent with
the boundary layer, from the stable component, which has a softer spectrum and
is consistent with the accretion disk.  At the HB/NB transition, the geometry
of the accretion flow appears to change, and we begin to lose the clear
distinction between the boundary layer and the accretion disk.  At the FB, the
thickening accretion disk may encompass the whole NS.  

In this model, as the geometry changes over the Z-track, the nature of the
reprocessed emission will also change.  As the source goes down the NB to the
FB and we lose sight of the boundary layer, the hard X-ray flux decreases.  At
the same time, we observe more of the thickening accretion disk, and as this is
the source of the reprocessed emission the NUV flux will increase.  This model
may also be consistent with the interpretation of the line fluxes by
\citet{shjny09}, and an increase of $\dot m$ over the Z-track.  Furthermore, in
this model the boundary layer is still visible in the HB/NB, in spite of the
large inclination angle.

As discussed above, the present observations are a challenge to the simple
interpretation of \citet{hvedm90} where $\dot m$ increases monotonically along
the Z-track of \cyg{} from HB-NB-FB. If changes in $\dot m$ lead directly to
changes the the temperature of the blackbody component, as may be suggested by
some models~\citep{ps01}, then it may be that $\dot m$ increases as the
hard color increases.  On the other hand, \citet{lrh09} and \citet{hvfrw10}
have suggested that Z sources have roughly constant $\dot m$, but that
different mechanisms, possibly related to the size of the boundary layer and
inner disk radius, are responsible for changing the spectrum along the
Z-track. The model of \cite{rg06}, with a thickening accretion disk that
encompasses the NS boundary layer, is nominally consistent with many of these
interpretations.  Without sufficient spectral coverage of the hard X-rays in
the 10-20 keV range to consistently track the blackbody component, our present
observations are insufficient to improve on the present uncertain state of
affairs.

The question of whether viewing geometry or a thickening accretion disk is the
source of the anticorrelation between NUV and hard X-ray flux is easily
testable with further observing campaigns of sources at different inclination
angles.  Unfortunately, most known NSXRB systems are
too extincted in the NUV for such a study.  However, other sources such as
4U~0614+09 have sufficiently low column density and large variability to be good
targets for follow-up work.

\acknowledgements
We thank \swift{} for support through the Guest Investigator program. We would
also like to sincerely thank the \swift{} operations team for their scheduling
of these observations. ESR thanks the TABASGO foundation.  EMC gratefully
acknowledges support provided by NASA through the {\it Chandra} Fellowship
Program, grant PF8-90052.  We also thank M. Revnivtsev and the anonymous
referee for helpful comments.


\end{document}